\documentclass[sigconf]{acmart}

\AtBeginDocument{%
  \providecommand\BibTeX{{%
    \normalfont B\kern-0.5em{\scshape i\kern-0.25em b}\kern-0.8em\TeX}}}

\setcopyright{acmlicensed}\acmConference[CHI '23]{Proceedings of the 2023 CHI Conference on Human Factors in Computing Systems}{April 23--28, 2023}{Hamburg, Germany}
\copyrightyear{2023}
\acmYear{2023}

\acmConference[CHI '23]{CHI '23: ACM Conference on Human Factors in Computing Systems}{April 23--April 28, 2023}{Hamburg, Germany}
\acmBooktitle{CHI '23: ACM Conference on Human Factors in Computing Systems, April 23--April 28, 2023, Hamburg, Germany}
\acmPrice{15.00}
\acmISBN{978-1-4503-XXXX-X/18/06}
\acmDOI{10.1145/3544548.3580969}
\acmISBN{978-1-4503-9421-5/23/04}

\usepackage[capitalize, noabbrev]{cleveref}
\usepackage{tabularx}
\usepackage{booktabs}
\usepackage{multirow}
\usepackage{tikz}
\usepackage{caption}
\usepackage{subcaption}
\usetikzlibrary{shadows}

\DeclareMathSymbol{\mhyphen}{\mathord}{AMSa}{"39}

\definecolor{changeNoteColor}{rgb}{0.1,0.6,1}
\newcommand{\changenote}[1]{#1} %

\begin{document}

\title[How Users Write with LLMs using Diegetic and Non-Diegetic Prompting]{Choice Over Control: How Users Write with Large Language Models using Diegetic and Non-Diegetic Prompting}

\author{Hai Dang}
\email{hai.dang@uni-bayreuth.de}
\orcid{0000-0003-3617-5657}
\affiliation{%
  \institution{University of Bayreuth}
  \streetaddress{Universitätsstr. 30}
  \city{Bayreuth}
  \state{Bavaria}
  \country{Germany}
  \postcode{95447}
}

\author{Sven Goller}
\email{sven.goller@uni-bayreuth.de}
\affiliation{%
  \institution{University of Bayreuth}
  \streetaddress{Universitätsstr. 30}
  \city{Bayreuth}
  \state{Bavaria}
  \country{Germany}
  \postcode{95447}
}

\author{Florian Lehmann}
\orcid{0000-0003-0201-867X}
\email{florian.lehmann@uni-bayreuth.de}
\affiliation{%
  \institution{University of Bayreuth}
  \streetaddress{Universitätsstr. 30}
  \city{Bayreuth}
  \state{Bavaria}
  \country{Germany}
  \postcode{95447}
}

\author{Daniel Buschek}
\orcid{0000-0002-0013-715X}
\email{daniel.buschek@uni-bayreuth.de}
\affiliation{%
  \institution{University of Bayreuth}
  \streetaddress{Universitätsstr. 30}
  \city{Bayreuth}
  \state{Bavaria}
  \country{Germany}
  \postcode{95447}
}

\renewcommand{\shortauthors}{Dang et al.}

\newcommand{\lastaccessed}{\textit{last accessed \today}}

\definecolor{HaisColor}{rgb}{0.9,0.3,0.9}
\newcommand{\hai}[1]{\textsf{\textbf{\textcolor{HaisColor}{[Hai: #1]}}}}

\definecolor{DanielsColor}{rgb}{0.9,0.6,0.1}
\newcommand{\daniel}[1]{\textsf{\textbf{\textcolor{DanielsColor}{[Daniel: #1]}}}}

\definecolor{SvensColor}{rgb}{0.5,0.8,0.5}
\newcommand{\sven}[1]{\textsf{\textbf{\textcolor{SvensColor}{[Sven: #1]}}}}

\definecolor{deemphColor}{rgb}{0.4,0.4,0.4}
\newcommand{\deemph}[1]{\textcolor{deemphColor}{#1}}

\newcommand{\pct}[1]{#1\,\%} %
\newcommand{\ms}[1]{#1\,ms}
\newcommand{\secs}[1]{#1\,s}
\newcommand{\mins}[1]{#1\,min}

\newcommand{\mSD}[2]{(mean = #1, SD = #2)}
\newcommand{\cor}[4][p]{($r_#1$(#2) = #3, p = #4)}
\newcommand{\glmmci}[5]{$\beta$=#1, SE=#2, CI$_{95\%}$=[#3, #4], p#5}
\newcommand{\ttest}[7]{mean diff. = #1, CI$_{95\%}$=[#2, #3], t(#4) = #5, p#6; d = #7}
\newcommand{\ttestcohend}[3]{t=#1, p#2, d=#3}

\newcommand{\ivinstruction}{\textsc{Instruction}}
\newcommand{\ivnumber}{\textsc{Number}}

\newcommand{\qx}[1]{$Q_{#1}$}

\newcommand{\suggnone}{s_0}
\newcommand{\suggone}{s_1}
\newcommand{\suggthree}{s_3}

\newcommand{\instructyes}{i_{yes}}
\newcommand{\instructno}{i_{no}}

\newcommand{\prestudyN}{6}
\newcommand{\mainStudyNraw}{X}
\newcommand{\mainStudyN}{129}
\newcommand{\mainStudyM}{71}
\newcommand{\mainStudyF}{57}
\newcommand{\mainStudyNB}{1}

\newcommand{\participant}[1]{$P_{#1}$}
\newcommand{\coded}[1]{(coded #1 times)}

\newcommand*\keystroke[1]{%
  \tikz[baseline=(key.base)]
    \node[%
      draw,
      fill=white,
      drop shadow={shadow xshift=0.25ex,shadow yshift=-0.25ex,fill=black,opacity=0.75},
      rectangle,
      rounded corners=2pt,
      inner sep=1pt,
      line width=0.5pt,
      font=\scriptsize\sffamily
    ](key) {#1\strut}
  ;
}

\begin{abstract}
  
We propose a conceptual perspective on prompts for Large Language Models (LLMs) that distinguishes between (1) diegetic prompts (part of the narrative, e.g. \textit{``Once upon a time, I saw a fox ...''}), and (2) non-diegetic prompts (external, e.g. \textit{``Write about the adventures of the fox.''}).
With this lens, we study how \mainStudyN{} crowd workers on \textit{Prolific} write short texts with different user interfaces (1 vs 3 suggestions, with/out non-diegetic prompts; implemented with \textit{GPT-3}): 
When the interface offered multiple suggestions and provided an option for non-diegetic prompting, participants preferred choosing from multiple suggestions over controlling them via non-diegetic prompts. When participants provided non-diegetic prompts it was to ask for inspiration, topics or facts.
Single suggestions in particular were guided both with diegetic and non-diegetic information. 
This work informs human-AI interaction with generative models by revealing that (1) writing non-diegetic prompts requires effort, (2) people combine diegetic and non-diegetic prompting, and (3) they use their draft (i.e. diegetic information) and suggestion timing to strategically guide LLMs.

\end{abstract}

\ccsdesc[500]{Human-centered computing~Empirical studies in HCI}
\ccsdesc[500]{Human-centered computing~Text input}
\ccsdesc[500]{Computing methodologies~Natural language generation}

\keywords{Large language models, Co-creative systems, Human-AI collaboration, User-centric natural language generation}

\begin{teaserfigure}
 \includegraphics[width=\textwidth]{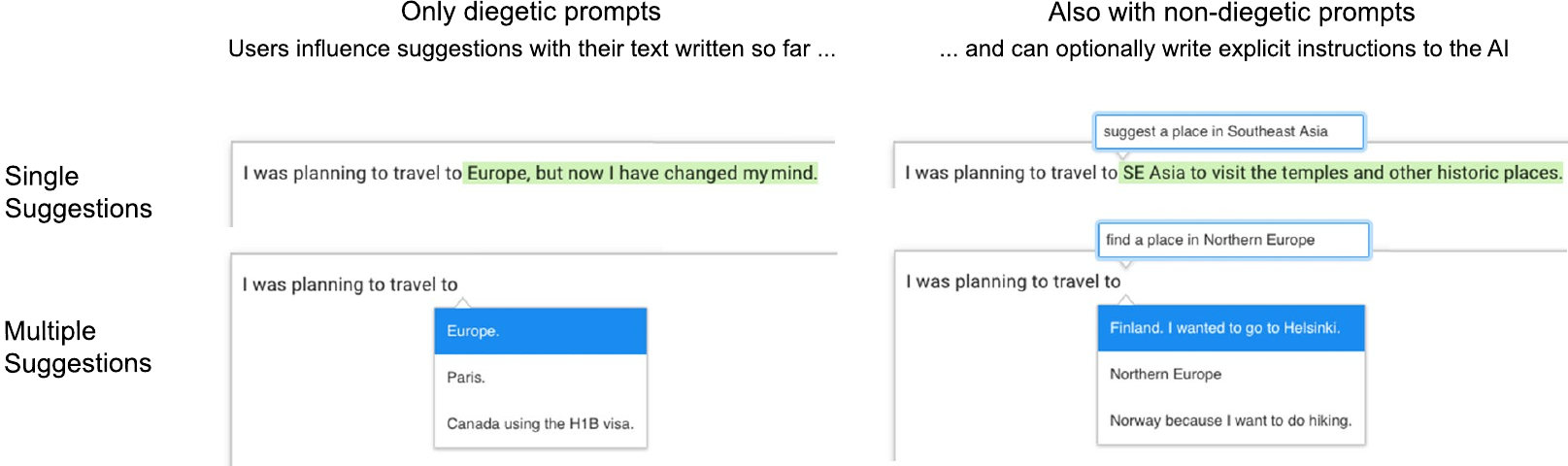}
 \caption{Overview of our four UI variants, showing the user's written text (black font, i.e. a diegetic prompt), the suggestions (text highlighted in green, and options in the list), and a popup text box that allows users to input an instruction as a zero-shot prompt to the system (i.e. a non-diegetic prompt).}
 \Description{A screenshot showing four writing setups used in the study. The setups are categorized in: Single vs. Multiple Suggestions and Only Diegetic vs. Also Non-Diegetic Prompts. Users written text document represents the diegetic prompt whereas the explicit instruction to the AI is the non-diegetic prompt.}
 \label{fig:teaser}
\end{teaserfigure}

\maketitle

\section{Introduction}

When writing collaboratively, people coordinate and inspire each other through what they write in the draft itself and through communication beyond it. In this paper, we examine related mechanisms for human-AI co-writing.

Input text provided to a Large Language Model (LLM) as a basis for generating text is referred to as a ``prompt''. Providing a few examples of inputs and outputs in such a text prompt can help the model solve a task \cite{Brown2020gpt3, zhao_calibrate_2021}. This is called few-shot learning. For example, an LLM can be prompted to translate from English to French with a few examples of English sentences and corresponding translations, followed by the English sentence to be translated. By completing this text the LLM then (ideally) translates that sentence. This affords user control: Users can define tasks and delegate them to an LLM ad-hoc.
Going further, zero-shot learning prompts the LLM with an instruction \textit{without} examples (e.g. \textit{Translate 'The weather is nice' to French}). This is a harder task but from a Human-Computer Interaction (HCI) point of view it frees users from thinking of specific examples when instructing the AI system. %

We introduce the terms diegetic prompting and non-diegetic prompting\footnote{\url{https://www.merriam-webster.com/dictionary/diegetic}, \lastaccessed} to frame a new perspective on how users influence an LLM in their writing process. A diegetic prompt is part of the users' narrative. For example, when the user writes about a vacation in South East Asia, the story as written so far forms the diegetic prompt. In contrast, a non-diegetic prompt is an explicit instruction to the LLM (e.g. ``suggest activities to do in Singapore''). Crucially, this instruction is not a part of the resulting document (e.g. travel blog); it only serves to guide the LLM's text generation. %

Technically, there may not be a difference between diegetic and non-diegetic prompts for the LLM -- both types are received by the model as text input strings. However, from an HCI perspective, this distinction allows us to identify patterns in the perception and interaction of users writing with LLMs.
With this new distinction, in this paper we address the research question: \textit{How do users write with Large Language Models using diegetic and non-diegetic prompting?}

Concretely, we propose and compare four UI variants (\cref{fig:teaser}) that allow people to write with these types of prompts, plus a baseline UI without suggestions. We conducted a remote study with \mainStudyN{} crowd workers on Prolific, each writing five stories. We investigate the influence of two independent variables on users' writing behavior, namely \ivinstruction{} with two levels ($\instructno$, $\instructyes$) and \ivnumber{} of suggestions with three levels (baseline: $\suggnone$; $\suggone$, $\suggthree$).

Users overall prefer choosing from multiple suggestions over controlling them via non-diegetic prompts. They use non-diegetic prompts to ask the LLM for inspiration, topics or facts. Non-diegetic prompts increase effort, for learning how to formulate them and switching between diegetic and non-diegetic writing. 
Users also prefer the UI with multiple suggestions over seeing single ones, yet allowing them to provide non-diegetic prompts reduces the gap in acceptance rates by boosting it for the UI with single suggestions.
Moreover, single suggestions are triggered later in sentences, and less frequently at transition words and to start sentences. Together with people's comments, this indicates that writers consider diegetic information to guide LLMs. %
We discuss implications for LLMs and interaction design.

We contribute a new conceptual lens on prompting that distinguishes diegetic and non-diegetic ways in which users can influence LLMs, and a new UI design to combine text continuation suggestions with zero-shot prompt input.%

\section{Related Work}

We relate our work to prompting in Natural Language Processing (NLP) and writing interfaces in Human-Computer Interaction (HCI). Moreover, we present our proposed concept of diegetic and non-diegetic prompting by locating it in existing user interfaces for writing and prompting.

\subsection{Prompting in Large Language Models}
Language models are trained to predict the next word given the previous words in the text. One primary advantage of Deep Learning-based LLMs is that they can solve several natural language processing tasks without being specifically trained on those. This can be done via text prompts written in natural language \cite{Brown2020gpt3}. \citet{zhao_calibrate_2021} show that providing a few examples of inputs and outputs can help to steer the model. However, optimizing prompts is not trivial and requires extensive experience \cite{liu_pre_train_2021}.

\subsubsection{Prompt Engineering}
Related work in prompt engineering has proposed several methods to improve prompts: For example, paraphrasing prompts can lead to better model outputs \cite{jiang_how_2020, yuan_bartscore_2021, liu_what_2021, haviv_bertese_2021}. Another approach involves constructing prompt templates to increase the accuracy for probing knowlege \cite{petroni_language_2019}, for translation tasks \cite{Brown2020gpt3}, or for text classification tasks \cite{schick_exploiting_2021}. However, optimized prompts constructed in the process of prompt engineering are usually not meant to be consumed by humans; rather, they are designed for LLMs to most effectively perform a task \cite{liu_pre_train_2021}. In contrast, in our study, we explore how non-expert users write and use (zero-shot) prompts when writing with an LLM.

\subsubsection{Prompting Interfaces}
Several interactive systems have been proposed to enable users to work more effectively with prompts: For example, \textit{AI Chains} by \citet{wu_aichains_2022} allows users to combine multiple prompt primitives and their outputs to form a chain of prompts that can solve complex language processing tasks. In another study, they introduce an interface for visually programming these chains \cite{wu_promptchainer_2022}. Similarily, \textit{PromptMaker} \cite{jiang_promptmaker_2022} allows users to prototype new AI functionalities using language prompts. \citet{strobelt_interactive_2022} developed a prompt programming environment to allow users to experiment with prompt variations and visualize prompt performance. \textit{Story Centaur} by \citet{swanson_story_centaur_2021} supports users in creating few-shot examples for creative writing. Using our terminology, these projects focused on \textit{non-diegetic} prompts as a main output of interaction. In contrast, we integrate non-diegetic prompts into a text editor, with a focus on writing.
Concretely, we combine a UI for phrase suggestions with a UI for zero-shot prompt inputs to an LLM, and analyze how users make use of these during their writing.

\subsection{Writing Interfaces for LLMs}\label{sec:related_work_writing_uis}

Here we give a brief overview of key design factors for user interfaces that involve LLMs and text generation.

\subsubsection{Scope of Suggestions}
Earlier work mainly focused on single word suggestions \cite{Dunlop2012, Fowler2015, Gordon2016, Quinn2016chi}. This scope favours performance metrics, such as reducing key-strokes, while longer phrase suggestions \cite{roemmele_creative_2015, buschek_2021, Lee_coauthor_2022} are perceived more as new ideas for writing~\cite{Arnold2018}. We focus on such phrase suggestions in this paper. %

\subsubsection{Display of Suggestions}
Single text suggestions can be shown inline~\cite{Bhat2022, Yuan_wordcraft_2022, Chen2019, goodman_lampost_2022}, whereas multiple suggestions are shown as pop-up lists of about three to six entries~\cite{buschek_2021, Lee_coauthor_2022}. Beyond that, \citet{Singh_2022} evaluated how writers use suggestions displayed as images and sound. Moreover, \citet{Bhat2022} used a a pop-up text box to show suggestions for insertions in the middle of sentences. We follow these design choices (\cref{fig:teaser}) and show single suggestions inline and multiple ones in a pop-up list. We add a pop-up text field for entering non-diegetic prompts.

\subsubsection{Implicit vs. Explicit Trigger}
In writing interfaces, suggestions can be triggered explicitly or implicitly. Related work showed suggestions automatically after short inactivity \cite{Bhat2022, buschek_2021} or gated by a utility function \cite{Kannan_2016}. %
Alternatively, recent work has also explored designs in which users explicitly request suggestions with a hotkey \cite{calderwood_how_2020, Lee_coauthor_2022, Yuan_wordcraft_2022, Singh_2022, goodman_lampost_2022}. We also use this design with an explicit request key to better understand how and when users request suggestions.

\subsection{Diegetic and Non-Diegetic Prompting in Existing Writing Interfaces}
Here we apply the proposed lens to analyse how existing systems use diegetic and non-diegetic information in their writing interfaces. 
Traditionally, systems mainly use diegetic information, that is, they predict text based (only) on the preceding text \cite{ Dunlop2012, Fowler2015, Gordon2016, Quinn2016chi}. Some also added other information (e.g. hand posture, body movement~\cite{Goel_contexttype_2013, Goel_walktype_2012}). These show early examples of non-diegetic input to the language model. In this work, we focus on textual diegetic and non-diegetic information.

From a technical perspective, for recent systems that use LLMs to generate text suggestions, there might be no difference between the user's text draft (i.e. diegetic text) and other text inputs to the language model (e.g. instructions to the model, i.e. non-diegetic text). Therefore, systems in which the UI did not afford text prompts explicitly made the implicit choice of only using diegetic information as their input to the LLM \cite{buschek_2021, Lee_coauthor_2022, Singh_2022, calderwood_how_2020}. %

In contrast, writing interfaces that indeed allow users to explicitly enter prompts often use a mix of diegetic and non-diegetic information. \citet{gero_sparks_2022} propose ``sparks'', i.e. sentences generated from LLMs to inspire new ideas for scientific writing. The user-provided prompts to generate these sparks are not part of the final outcome text, thus they are non-diegetic. Similarly, other systems (e.g. \textit{Wordcraft}~\cite{Yuan_wordcraft_2022}, \textit{LaMPost}~\cite{ goodman_lampost_2022}) allowed users to select a part of the written text and modify it via predefined functionality (internally these functionalities also use prompting: e.g. a button for ``rewrite selection'' + text entry field for prompt). The selected text in this example is diegetic information while the prompt template and user-provided prompts are non-diegetic information. Related, we include the entire user written text draft as diegetic information and allow users to provide non-diegetic custom text prompts to further guide the LLM.

\section{Interaction Concept}\label{sec:concept}

Here we describe our UI and interaction concept (also see \cref{fig:teaser} and \cref{fig:screenshot_system}): It closely integrates diegetic and non-diegetic prompting in the same UI; users can use both types without having to take the hands off the keyboard.

\subsection{Inline (Single) Suggestions (\cref{fig:teaser} top row)}
When a user requests a new suggestion (\keystroke{TAB}) a preview of the suggestion appears after the current caret position in the text editor. Users can press \keystroke{TAB} repeatedly to get new suggestions. The suggestion preview is visually highlighted in green to indicate that it is not part of the text yet. We decided for this design instead of e.g. a greyed out suggestion text (as e.g. used in Google's Smart Compose~\cite{Chen2019}) because pilot tests showed that grey text can be difficult to read for some people and makes readability more dependant on screen brightness settings, which we cannot control in an online study. If the suggestion is accepted (\keystroke{ENTER}) the preview style (green background) is removed and the suggested text becomes part of the text document. Alternatively, the user can cancel the current suggestion preview by pressing \keystroke{ESC} or by continuing to type without confirming the suggestion. When the suggestion is cancelled in one of these ways, the previewed suggestion is removed from the text editor.

\subsection{Multiple Suggestions (\cref{fig:teaser} bottom row)}
Our system follows current practices for multiple suggestions (see \cref{sec:related_work_writing_uis}) and shows each phrase suggestion as a separate item in a list of three. Again, users can press \keystroke{TAB} to get suggestions (and repeatedly to get new ones). Users can use the \keystroke{$\uparrow$ UP/ $\downarrow$ DOWN} keys to navigate this list and confirm a suggestion with \keystroke{ENTER}. Selection via mouse is also possible. %

\subsection{Pop-up Textbox for Non-Diegetic Prompts (\cref{fig:teaser} right column)}\label{sec:concept_popup_textbox}

In the study, we described non-diegetic prompts as ``instructions to the AI''. Users can request suggestions with \keystroke{TAB} as before. Additionally, they can enter an instruction by typing in a popup box that appears above the caret position. Thus, users have the option to input an instruction but are not forced to do so to request suggestions. Input focus is automatically switched from the text editor to the pop-up textbox when requesting suggestions so users can type instructions directly after pressing \keystroke{TAB}. Users can submit the instruction with \keystroke{TAB} or \keystroke{Enter}. 
They can then press \keystroke{Enter} again to accept the selected suggestion. Alternatively, they can revise their instruction to update the suggestions.

\section{Prototype Implementation}\label{sec:implementation}
Here we provide details about the web prototype used in the study. %
For screenshots, see \cref{fig:teaser} and \cref{sec:appendix}.

\subsection{Web System}
The prototype was implemented with ReactJS\footnote{\url{https://reactjs.org/}, \lastaccessed} and CKEditor5\footnote{\url{https://ckeditor.com}, \lastaccessed}. Each suggestion request from the client was passed to and parsed by a backend server which used FastApi\footnote{\url{https://fastapi.tiangolo.com}, \lastaccessed} as a lightweight webserver. The server forwarded these requests to OpenAI's \textit{text-davinci-edit-001} model along with the entire written text as well as an instruction for the suggestion model (see \cref{sec:concept}). We chose this model and API because it is reportedly trained specifically to take in a given text as well as a (separate) instruction relating to the text.

\subsection{Language Model Prompts}
We used two default prompt prefixes to retrieve sentence completions from GPT-3 (text-davinci-edit-001): (1) \textit{Complete the sentence.} (2) \textit{Complete the sentence and <user\_instruction>.} The system automatically used (1) when there was no option for the participants to provide explicit instructions to the AI, or when users did not provide an instruction. When they did write instructions, these were appended to (2). For instance, if the user wrote the instruction: 'suggest colors', the resulting full instruction sent to the model was: 'Complete the sentence and suggest colors'. During the pre-study we experimented with other default instructions such as: 'Continue' or 'Continue the text', as well as more complex ones, but found them to be less suitable (e.g. produced longer text or less consistent). We applied a post processing step to trim the model's output and display only the generated continuation.

\subsection{Information Box}
For the user study, we implemented an information box (\cref{fig:screenshot_system} in \cref{sec:appendix}) which explains the different features of the current text editor setup. Concretely, it showed an image that demonstrates the usage of the UI as well as an explanation of the available action keys. 
\section{Method}\label{sec:method}

\changenote{We used the following methods, in line with related studies on human-AI writing (e.g. cf.~\cite{buschek_2021, Lee_coauthor_2022}).}

\subsection{Questionnaires}\label{sec:method_questionnaires}

\changenote{To assess participants' backgrounds,} an initial questionnaire asked about demographics and experience with writing features and language models. %
Participants also filled in one questionnaire after each UI variant (see \cref{fig:likert_agreements}) \changenote{to give subjective feedback per UI.} %
\changenote{To extend on this with overall feedback,} a final questionnaire asked for (optional) open comments on changes to the system and experiences with suggestions and instructions.

\subsection{Interaction Logging}

\changenote{To analyze interaction behaviour in detail,} we logged interaction events\changenote{, i.e. key and mouse events,} during the writing tasks (\changenote{see \cref{tab:event_log}}). Each event included a \textit{timestamp}, \textit{task id}, and the \textit{current text} in the editor. Depending on the event it included information such as the suggestion trigger position in the text or the instruction to the AI.

\subsection{Coding of Open Questions}
We analysed the open comments from the final questionnaire in an approach adopting coding steps from Grounded Theory~\cite{corbin1990basics, muller_grounded_2012}, \changenote{in order to identify and report on the emerging aspects}: 
First, two researchers inductively proposed codes for the data of 20 people. 
They then compared and clustered these codes to develop a common codebook. 
Then, they coded the first 20 plus 32 more participants and checked each other's codings, with slight adjustments to the codebook. Finally, one researcher coded the remaining data and another one checked this coding. Throughout the process, disagreements were resolved via discussion.

\subsection{\changenote{Evaluation of User Written Text}}

\changenote{We used LanguageTool\footnote{\url{https://languagetool.org}, \lastaccessed}, a multilingual grammar and spell-checker, to count the number of grammar and spelling mistakes. To evaluate the degree to which participants engaged with the selected writing prompts during the user study (cf. \cref{sec:user_study}), three researchers independently reviewed the stories and provided comments on their connection to the prompts. Finally, one researcher reviewed all comments to ensure consistency.}
\section{User Study}\label{sec:user_study}

\subsection{Study Design}\label{sec:study_design}
Our study uses a within-subject design with two independent variables: The \ivnumber{} of (parallel) suggestions with two levels: one and three suggestions ($\suggone$, $\suggthree$); and the opportunity for \ivinstruction{} with two levels ($\instructno$, $\instructyes$). This results in four UI variants with suggestions. In addition, we included a baseline UI without any suggestions ($\suggnone$). \changenote{The order of these five UIs was fully counterbalanced.} As dependent variables we included interaction measures as well as questionnaire data.

\subsection{Participants}
We conducted a pre-study with \prestudyN~participants with direct disussions for rich feedback, followed by our main study with \mainStudyN~participants (M=\mainStudyM , F=\mainStudyF , NB=\mainStudyNB ). We recruited on \textit{Prolific}\footnote{\url{https://www.prolific.co/}, \lastaccessed} and screened participants for written and spoken fluency in English, as well as access to a computer with a keyboard. Participants reported ages ranged from 18 to 70 with a median age of 32. Following the platform recommendations, participants were compensated with 8\,\pounds/h.

\subsection{Procedure}
The study started with a description page, including information about the collected data and GDPR, in line with our institute's regulations. After giving their consent, participants were directed to a page with an overview of the procedure and involved UI variants. Following this, people were guided through the five writing tasks in a counterbalanced order, and then to the final questionnaire. The study had an estimated duration of 45 minutes \changenote{(actual mean was 45 minutes and 41 seconds)}.  %

\subsubsection{Topic Selection}
For each task, participants first selected a writing topic. Repeated selections were allowed, as in related work~\cite{Lee_coauthor_2022}, yet we asked them to choose at least two different topics overall. %
The topic order was also randomized and shown one at a time to encourage variety in the topic choices overall.

\citet{gero_design_2022} suggested three tasks for writing support tools, including story writing and argumentative essay writing. We thus selected five topics for creative writing\footnote{\url{https://www.reddit.com/r/WritingPrompts}, \lastaccessed} and five topics for argumentative writing from the same source\footnote{\url{https://www.nytimes.com/2021/02/01/learning/300-questions-and-images-to-inspire-argument-writing.html}, \lastaccessed} as \citet{Lee_coauthor_2022}.  %

\subsubsection{Writing Task}
Participants were told to write about the previously selected topic for five minutes and finish their text with a clear ending. The description encouraged to try out all features but also to write their own text. A timer was shown below the text editor. It was mentioned that the timer was not a hard cut-off, but served as a reminder of when to finish the task. %
We set a minimum time of 15 seconds before participants could submit their story but people stayed close to the five minutes anyway (see \cref{sec:task_times}). People filled in a questionnaire after each task (\cref{sec:method_questionnaires}).

\section{Results}

Here we present our study results. For statistical testing we use R~\cite{R2020}, concretely, (generalised) linear mixed-effects models (LMMs with the packages \textit{lme4}~\cite{Bates2015}, \textit{lmerTest}~\cite{Kuznetsova2017}). The models account for participants' individual differences, as well as for the type of their chosen topics (creative story writing, argumentative writing), via random intercepts. As fixed effects, the models have \ivinstruction{} and \ivnumber{}.
Moreover, we use the R package \textit{multgee}~\cite{multgee2015} to analyse the Likert results (i.e. ordinal data) with Generalized Estimating Equations (GEEs). 
We report significance at p~<~0.05.

\begin{figure*}[t]
    \centering
    \includegraphics[width=\textwidth]{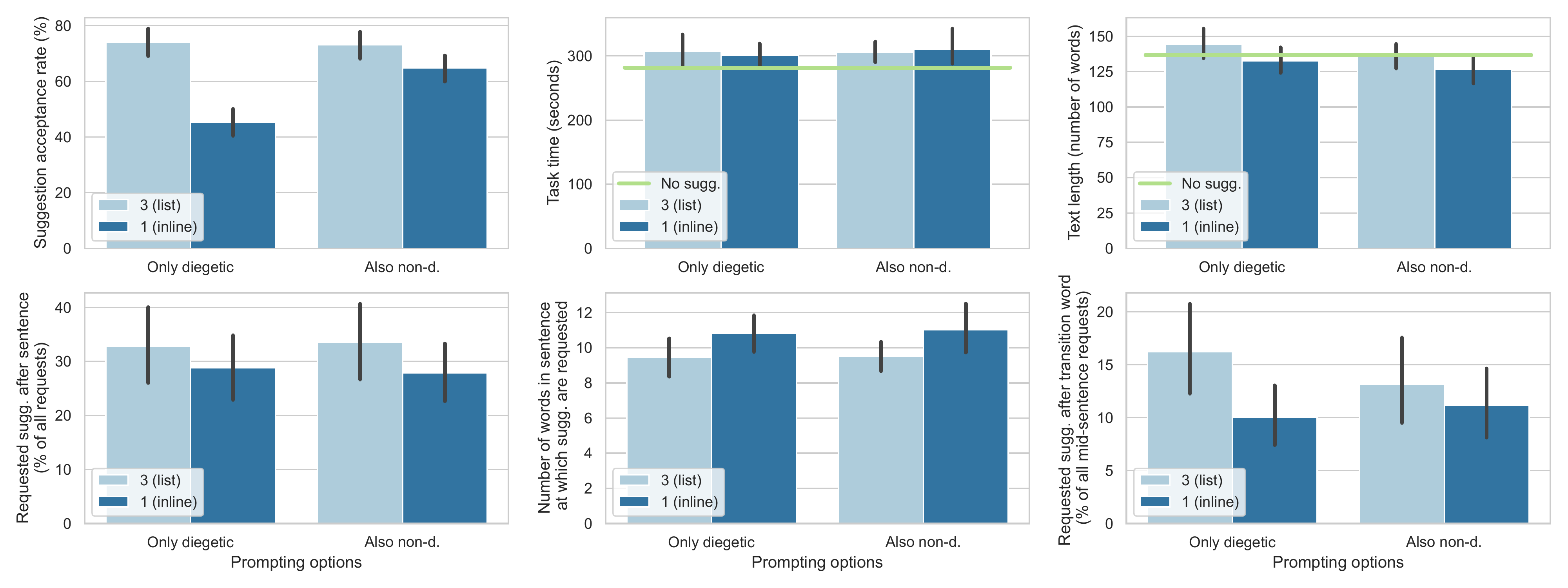}
    \caption{Overview of the interaction metrics in our study. In summary, we observe: (1) Giving users the option to write instructions (i.e. non-diegetic prompts) increases the acceptance rate of suggestions for single suggestions, but not beyond that of multiple suggestions (top left). (2) Writing time did not vary much and texts were slightly shorter with single suggestions and instructions (top center/right). (3) Single suggestions were requested less often at the start of sentences (bottom left), about 1.5 words later in a sentence (bottom center), and less often after transition words (bottom right). See text for details.}
    \label{fig:analysis_overview}
    \Description{Six bar charts showing: (1) The overall acceptance rate, (2) the task time, (3) the average text length of the submitted stories, (4) whether users requested suggestions at the start of a new sentence, (5) the number of words written within the current sentence before users trigger suggestions, and (6) whether users have requested suggestions after transition words.}
\end{figure*}

We define a \textit{suggestion session} as continuous interaction with suggestions, from requesting them until cancellation or acceptance (e.g. a session might involve three subsequent ``tab'' presses to browse suggestions). 
\changenote{Participants} triggered 3097 suggestion sessions. %
The mean in tasks with suggestions enabled was 6.47 (SD 4.00), comparable to related work~\cite{Lee_coauthor_2022}. %

\begin{table*}[]
\centering
\footnotesize
\newcolumntype{L}{>{\raggedright\arraybackslash}X}
\newcolumntype{P}[1]{>{\raggedright\arraybackslash}p{#1}}
\renewcommand{\arraystretch}{1.4}
\setlength{\tabcolsep}{4pt}
\begin{tabularx}{\linewidth}{lP{2.75em}P{8em}P{8.5em}P{10em}P{7em}L}
\toprule
  &
  \textbf{Section} &
  \textbf{Aspect} &
  \textbf{Sig. pos. predictors} &
  \textbf{Sig. neg. predictors} &
  \textbf{Sig. interaction} &
  \textbf{Takeaway in words} \\ \midrule
1 &
  \ref{sec:acceptance_rates} &
  Suggestion acceptance &
   &
  $\suggone$ \deemph{(\glmmci{-1.26}{0.11}{-1.48}{-1.03}{<.0001})} &
   \ivnumber{} * \ivinstruction{} \deemph{(\glmmci{0.72}{0.17}{0.39}{1.05}{<.0001})} &
  Showing one suggestion (instead of three) decreases chance of acceptance; more so without instructions than with them. \\
  \midrule 
2 &
  \ref{sec:task_times} &
  Task time, comparing sugg. UIs &
   &
   &
   &
  No sig. differences in task completion times were found between the four UIs with suggestions. \\
3 &
  \ref{sec:task_times} &
  Task time, comparing sugg. UIs against baseline (no suggestions) &
  $\suggthree$ \deemph{(\glmmci{24.59}{9.5}{5.96}{43.22}{<.01})} &
   &
   &
  Writing with three suggestions took longer than without suggestions. \\
  \midrule 
4 &
  \ref{sec:text_length} &
  Text length, comparing sugg. UIs &
   &
  $\suggone$ \deemph{(\glmmci{-0.08}{0.01}{-0.10}{-0.06}{<.0001})}; $\instructyes$ \deemph{(\glmmci{-0.06}{0.01}{-0.09}{-0.04}{<.0001})} &
   &
   Texts are slightly shorter when writing with single suggestions or with a UI allowing for instructions...
   \\
5 &
  \ref{sec:text_length} &
  Text length, comparing sugg. UIs against baseline (no suggestions) &
  $\suggthree$ \deemph{(\glmmci{0.05}{0.01}{0.03}{0.07}{<.0001})} &
  $\suggone$ \deemph{(\glmmci{-0.03}{0.01}{-0.05}{-0.01}{<.005})}; $\instructyes$ \deemph{(\glmmci{-0.05}{0.01}{-0.07}{-0.03}{<.0001})} &
   &
  ..., also compared to the baseline. However, writing with multiple suggestions leads to slightly longer texts. \\%
  \midrule 
6 &
  \ref{sec:sugg_moments_start_sentence} &
  Requesting suggestions after sentence vs mid-sentence &
   &
  $\suggone$ \deemph{(\glmmci{-0.83}{0.13}{-1.08}{-0.57}{<.0001})}; $\instructyes$ \deemph{(\glmmci{-0.32}{0.14}{-0.59}{-0.05}{=0.018})} &
   &
  Showing one suggestion (instead of three) decreased the chance of requesting suggestions at the beginning of a new sentence. \\
7 &
  \ref{sec:sugg_moments_num_words} &
  Number of words in sentence at suggestion request &
  $\suggone$ \deemph{(\glmmci{1.55}{0.78}{0.01}{3.08}{=0.049})} &
   &
   &
  Showing one suggestion (instead of three) increased the number of words in a sentence after which suggestions were requested. \\
8 &
  \ref{sec:sugg_moments_type_words} &
  Type of words at suggestion request &
   &
  $\suggone$ \deemph{(\glmmci{-0.37}{0.14}{-0.65}{-0.09}{=0.010})} &
   &
  Showing one suggestion (instead of three) decreased the chance of requesting suggestions after a transition word. \\ 
  \bottomrule
\end{tabularx}
\caption{\changenote{Overview of the (generalised) LMM results and takeaways of the significant results. Empty cells indicate no significant results. See Sections \ref{sec:acceptance_rates} - \ref{sec:sugg_moments} for details and \cref{fig:analysis_overview} for a descriptive overview of the data.}} %
\label{tab:lmm_overview}
\end{table*}

\subsection{Suggestion Acceptance}\label{sec:acceptance_rates}
We define the acceptance rate as the number of accepted suggestions divided by the number of triggered suggestion sessions. We found considerable differences between the UIs (Means: $\suggone$=0.55, $\suggthree$=0.74, $\instructno$=0.59, $\instructyes$=0.69). The grand mean acceptance rate was 0.64 (SD 0.29). The mean for suggestion requests with a written instruction was in line with this (0.64, SD: 0.33).
We fitted a generalised LMM on the acceptances as binomial data (i.e. for each shown suggestion we logged if it was accepted or not)\changenote{, summarized in \cref{tab:lmm_overview} (row 1). \cref{fig:analysis_overview} (top left) shows the descriptive data.}
In summary, showing one suggestion (instead of three) significantly decreased the chance of acceptance, yet enabling users to write instructions significantly reduced this gap by increasing their acceptance (Mean rate of 0.45 for $\suggone$ without instructions vs 0.65 with them).

\subsection{Task Completion Time}\label{sec:task_times}

We measured task time from starting the task to submitting it (Means:  $\suggnone$=281, $\suggone$=305, $\suggthree$=306, $\instructno$=303, $\instructyes$=309). \changenote{As a fixed writing time was given, we do not expect large differences here.}
\changenote{Indeed,} an LMM fitted on this data for the suggestion UIs did not reveal significant effects \changenote{(\cref{tab:lmm_overview}, row 2)}.
Another such model compared the suggestion UIs against the baseline \changenote{(\cref{tab:lmm_overview}, row 3)}: Here we found that writing with three suggestions took significantly longer than without suggestions.
This is in line with the descriptive picture in \cref{fig:analysis_overview} (top center): \changenote{Participants} followed the task description of writing for five minutes, and writing with the suggestion UIs took slightly longer. %

\subsection{Text Length}\label{sec:text_length}

In total, \changenote{submitted texts} contained 87,640 words, \changenote{including text from accepted suggestions}. The grand mean number of words per text was 134 words (SD 54).
We fitted a generalised (Poisson) LMM on the word count data to compare the four tasks with suggestions \changenote{(\cref{tab:lmm_overview}, row 4), and another such model to compare the suggestion UIs against the baseline without suggestions (\cref{tab:lmm_overview}, row 5)}.
The results match the descriptive pattern visible in \cref{fig:analysis_overview} (top right): In summary, texts are significantly shorter when writing with single suggestions or with a UI allowing for instructions. However, writing with multiple suggestions leads to significantly longer texts. %
These differences are rather small, about 6-10 words (Means: $\suggnone$=136, $\suggone$=130, $\suggthree$=140, $\instructno$=138, $\instructyes$=131).

\subsection{Moments of Suggestion Requests}\label{sec:sugg_moments}
We analysed at which moments \changenote{participants} requested suggestions. 

\subsubsection{After Sentence vs Mid-sentence}\label{sec:sugg_moments_start_sentence}
We analysed how often suggestions started a sentence (e.g. ``Hello, world! \textit{[tab]}'') vs in the middle (e.g. ``Hello world, how \textit{[tab]}'').
We fitted a generalised LMM on the requests as binomial data (i.e. for each request we logged if it was at the beginning of a new sentence or not), \changenote{summarized in \cref{tab:lmm_overview} (row 6). \cref{fig:analysis_overview} (bottom left) shows the descriptive data.}
In summary, showing one suggestion (instead of three) significantly decreased the chance of requesting suggestions at the beginning of a new sentence (Means: $\suggone$=\pct{21.70}, $\suggthree$=\pct{31.02}). %

\subsubsection{Number of Words in Sentence}\label{sec:sugg_moments_num_words}
For the suggestion requests in the middle of sentences we further analysed after how many words in that sentence they were requested.
We fitted an LMM on the mean numbers of words in sentences with suggestion requests per text, \changenote{summarized in \cref{tab:lmm_overview} (row 7). \cref{fig:analysis_overview} (bottom center) shows the descriptive data.} %
In summary, showing one suggestion (instead of three) significantly increased the number of words in a sentence after which suggestions were requested -- by about 1.5 words (Means: $\suggone$=10.93, $\suggthree$=9.48; i.e. a relative increase of \pct{15.3}), while \ivinstruction{} seemed to make no difference (Means: $\instructno$=10.18, $\instructyes$=10.34). \changenote{Note that 1-2 words later in a sentence is considerable because it may lead to very different constraints that users give to the system for possible continuations (e.g. ``The...'' vs ``The man said...')'.}

\subsubsection{Words at the Suggestion Requests}\label{sec:sugg_moments_type_words}
We further analysed the type of words after which suggestions were requested. Concretely, we categorised these ``trigger words'' into \textit{transition} words and other words, using online lists of English transition words\footnote{\raggedright e.g.: \url{https://www.grammarly.com/blog/transition-words-phrases/}, \url{https://writingcenter.unc.edu/tips-and-tools/transitions/}, \lastaccessed}. For example, transition words mark causes (e.g. ``because'', ``since''), opposites (e.g. ``while'', ``despite''), effects (e.g. ``therefore'', ``then''), and other aspects. We provide the full list we used in the project repository.
We fitted a generalised LMM on the requests as binomial data (i.e. for each request we logged if it was after a transition word or not), \changenote{summarized in \cref{tab:lmm_overview} (row 8). \cref{fig:analysis_overview} (bottom right) shows the descriptive data.} %
In summary, showing one suggestion (instead of three) significantly decreased the chance of requesting suggestions after a transition word (Means: $\suggone$=\pct{11.80}, $\suggthree$=\pct{14.03}), while \ivinstruction{} seemed to make no (sig.) difference (Means: $\instructno$=\pct{12.03}, $\instructyes$=\pct{13.32}).

\subsection{Perception of the Tasks and UIs}\label{sec:results_likert}

\begin{figure*}[t]
    \centering
    \includegraphics[width=\textwidth]{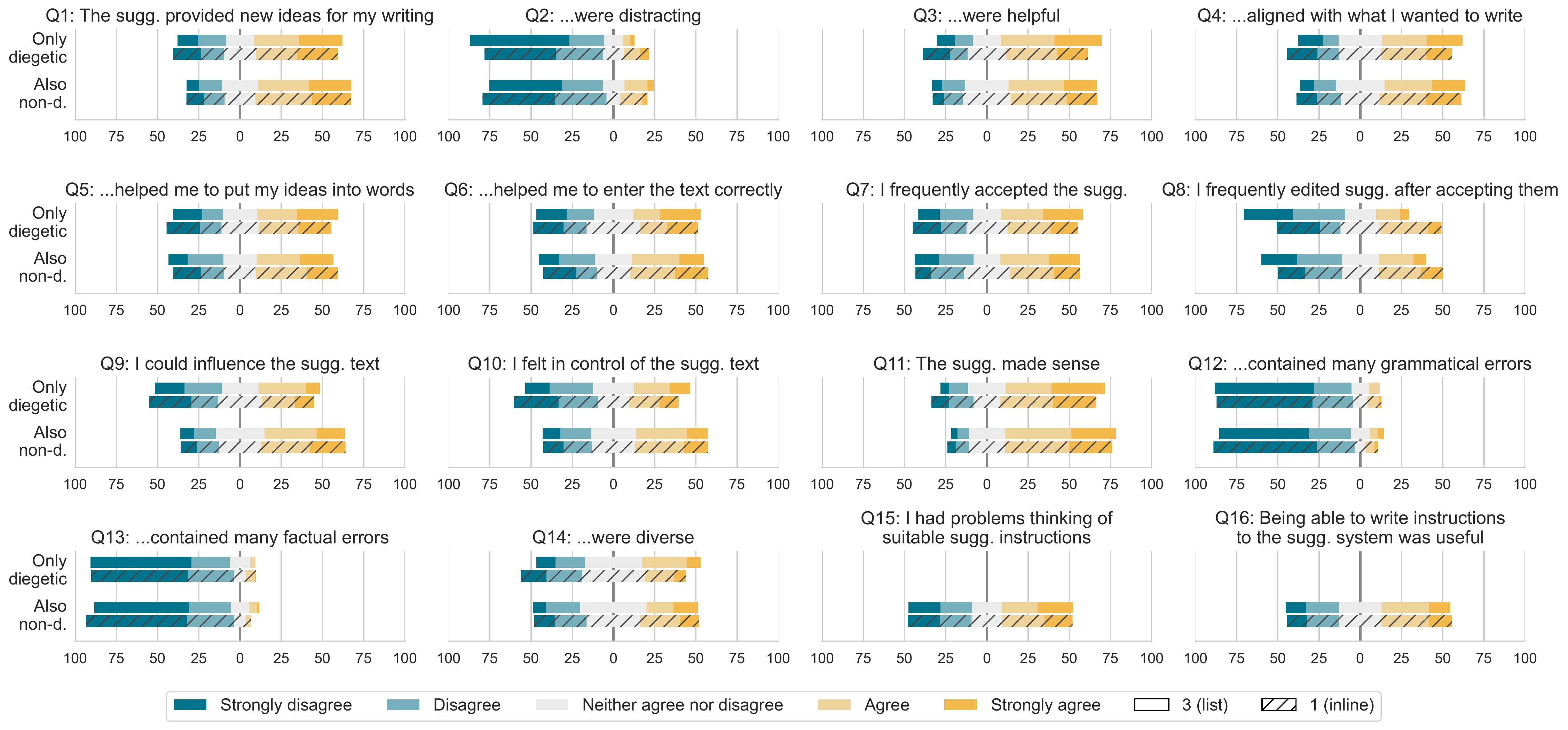}
    \caption{Overview of the Likert results. These questions were asked after each writing task. Note that Q15 and Q16 relate to the instructions (i.e. non-diegetic prompts) and thus were only asked for the corresponding tasks.}
    \label{fig:likert_agreements}
    \Description{Multiple likert scaled bar charts showing user responses for the between tasks questionnaires.}
\end{figure*}

We used Likert \changenote{items to assess participants'} perception after each writing task (\cref{fig:likert_agreements}).
Descriptively, suggestions received favourable ratings by the majority and had almost no perceived grammatical or factual errors. However, no UI was clearly ``best'' for everyone: Across questions and UI variants, there is a spread of opinions, including for the perceived usefulness of being able to write instructions (i.e. non-diegetic prompting). This spread fits to the different pros and cons and preferences that \changenote{participants} commented on (see \cref{sec:open_comments}).
Here, we report on the results from our GEE analysis. Since we have 16 questions, we summarize this analysis according to the emerging bigger picture.

\subsubsection{Perceived Differences for \ivnumber{} of Suggestions ($\suggone$ vs $\suggthree$)}
Showing a single suggestion was rated worse than having a list of three suggestions. This was significant for several questions. \changenote{The} GEE model \changenote{estimates} that the odds of giving a higher rating with a single suggestion were \changenote{``$x$'' times} the odds of that with the list of three suggestions, \changenote{with $x$ as follows}: Single suggestions were rated as significantly more distracting ($x$=1.95, p<0.005), less helpful ($x$=0.60, p=0.02), leading to more manual editing ($x$=1.91, p<0.005), feeling less in control ($x$=0.67, p=0.03), and providing less diverse suggestions ($x$=0.67, p=0.04).

\subsubsection{Perceived Differences for \ivinstruction{} ($\instructno$ vs $\instructyes$)}
We found a tradeoff in the perception of instructions: On the negative side, the UIs that allowed users to enter instructions to the AI received ratings of manually editing suggestions significantly more ($x$=1.54, p=0.02) and being significantly more distracting ($x$=2.03, p<0.0005). 

On the positive side, giving instructions was rated significantly better on being able to influence the suggested text ($x$=1.92, p<0.001). Descriptively, it was also rated better on feeling in control of the suggested text (see \qx{10} in \cref{fig:likert_agreements}), although this was not significant ($x$=1.42, p=0.058).

\subsubsection{Interactions of \ivnumber{} and \ivinstruction{}}

As mentioned in the previous two parts, both single suggestions and the ability to give instructions were perceived as significantly more distracting. However, there was also a significant negative interaction effect of \ivinstruction{} and \ivnumber{} on distraction. \changenote{The increase in distraction between $\suggone$ compared to $\suggthree$ was lower for $\instructyes$ than $\instructno$ (which also matches the picture for \qx{2} in \cref{fig:likert_agreements})}. This seems to be in line with the earlier finding for acceptance rates (\cref{sec:acceptance_rates}): Possibly, finding more useful single suggestions with instructions reduced the otherwise perceived distraction of single suggestions and/or instructions. That said, note that for all suggestion UIs, the majority did not find them distracting. We return to the aspect of distraction in more detail when analysing the open feedback (\cref{sec:open_comments}).

Finally, we also asked two questions that focused on the instructions directly (\qx{15} and \qx{16}) and thus could only be asked for those UIs with instructions (i.e. there's only a non-diegetic row in \cref{fig:likert_agreements} for \qx{15} and \qx{16}). For these two questions, we found no significant differences between $\suggone$ and $\suggthree$.

\subsection{Instruction Usage and Content}\label{sec:instruction_usage}

In total, participants used the non-diegetic prompting option to send 397 instructions to the system, with an average of 3.08 instructions per person (SD: 3.40).
The mean instruction length was 14.20 characters (SD: 9.01) and 2.52 words (SD: 1.74).
On average, \changenote{participants} had a ratio of 0.19 (SD: 0.17) of entering an instruction text when requesting suggestions, for those tasks that offered to do so. That is, about every fifth suggestion request used instructions.

We identified three main instruction ``styles'': The most common one (171 usages) was to use single \textit{keywords} (or comma-separated lists of keywords). We also found an \textit{imperative} style with 59 occurrences (e.g. starting the prompt text with ``suggest'', ``give'', ``find'', ``describe''). In 12 cases, \changenote{participants} formulated a \textit{question} (e.g. starting with a w-word like ``what'', ``who'' and so on, and/or ending with a ``?''). Other cases included instructions consisting of multiple words to describe something (e.g. ``somewhere in Italy'').
Qualitatively, we found a range of approaches (\cref{tab:instruction_examples}).

\begin{table}[t]
\centering
\footnotesize
\newcolumntype{L}{>{\raggedright\arraybackslash}X}
\renewcommand{\arraystretch}{1.2}
\begin{tabularx}{\linewidth}{p{10em}L}
\toprule
\textbf{Approach} &
  \textbf{Examples} \\ \midrule
providing a topic &
  ``school'', ``book'', ``retirement'', ``zoo'', ``event'' \\
providing adjectives &
  ``good'', ``horrendous”, ``bad”, ``long, too much, insane'', ``friendly'', ``funny'', ``scared'' \\
request for inspiration &
  ``give me a horror story'', ``suggest a place'', ``suggest the next step'', ``things we do in the morning'', ``suggest an activity for a middle aged man'' \\
make idea more concrete &
  ``suggest something disgusting'', ``what is wrong with dad'', ``suggest a cocktail'', ``suggest a type of pistol'' \\
request for variation &
  ``another phrase'', ``another action outside'', ``suggest a different approach'', ``anything'' \\
request for writing help &
  ``other words for stereotypical'', ``find a synonym for valued'', ``suggest a word for young people'', ``another word for talent'' \\
ask for opinion/advice &
  ``are books good'', ``what do i do next'' \\
retrieve facts &
  ``closest galaxy'', ``side effect of anti ageing'', ``a place on the Danube''\\
  \bottomrule
\end{tabularx}
\Description{Table shows examples for user provided instructions. }
\caption{An overview of the different approaches for writing non-diegetic prompts during the user study. Participants used single keywords to suggest topics and adjectives. Multiple-word prompts were often written in the imperative style, phrased as questions or phrased as incomplete sentences without a verb.}

\label{tab:instruction_examples}
\end{table}

\subsection{\changenote{Evaluation of Text Quality}}\label{sec:text_evaluation}
\changenote{The mean number of spelling and grammar mistakes per word was 0.0025, which is comparable to values reported in previous research \cite{Lee_coauthor_2022}. Approximately 3.5\%  of all texts (23 out of 645) did not align with the selected topics. Of these, the majority (13 out of 23) were written for the category of ``shapeshifter'', which may have been misunderstood as a metaphor for a specific set of desired traits in a partner. Despite this potential misunderstanding, the majority of participants demonstrated attentiveness to the task and provided thoughtful reflections on the topic.}

\section{Open Feedback}\label{sec:open_comments}

We analyzed the final feedback as described in \cref{sec:method}. We structure this report by the emerging aspects.

\subsection{Comments on Suggestions}

The majority preferred multiple suggestions (75 people stated this preference vs 23 for single suggestions). Main reasons were 
higher chances of finding fitting suggestions
\changenote{\coded{36} %
} and more inspiration 
\changenote{
\coded{8}}. %
As \changenote{\participant{38}} wrote: \textit{``I found the multiple suggestions much more user friendly and also much more inspiring due to the multiple options.''}

Those preferring single suggestions found them more intuitive
\changenote{\coded7}, %
faster to work with 
\changenote{\coded3} %
or less distracting \changenote{\coded3}: %
\textit{``I strongly preferred inline due to how intuitive they were to use.''} \changenote{(\participant{113})} Or: \textit{``I like seeing how the sentences actually looks in its actual place, and the inline suggestions allowed this.''} \changenote{(\participant{109})}.
Others liked not having to decide
\changenote{\coded2} %
but noted that this might lead to choosing a less than optimal suggestion.

Fourteen \changenote{participants} reflected on benefits for both, such as: \textit{``On the one hand, the inline suggestions felt less cluttered and I could just press tab again if the first suggestion wasn't suitable. On the other hand, displaying multiple suggestions at once could lead me to a better suggestion when I might just have settled for the first one.''} \changenote{(\participant{13})}

\changenote{Participants} commented on why and how to use suggestions, mentioning inspiration
\changenote{\coded{39}}, %
overcoming writer's block
\changenote{\coded{3}}, %
or finishing sentences
\changenote{\coded7}. %
For example: \textit{``I used the suggestions if they aligned with what I was writing or if I felt a little stuck with what to say next.''} \changenote{(\participant{117})} Or: \textit{``I tended to start with a vague idea of my own and see what ideas it had.''} \changenote{(\participant{14})}.

\begin{figure*}
    \centering
    \includegraphics[width=\textwidth]{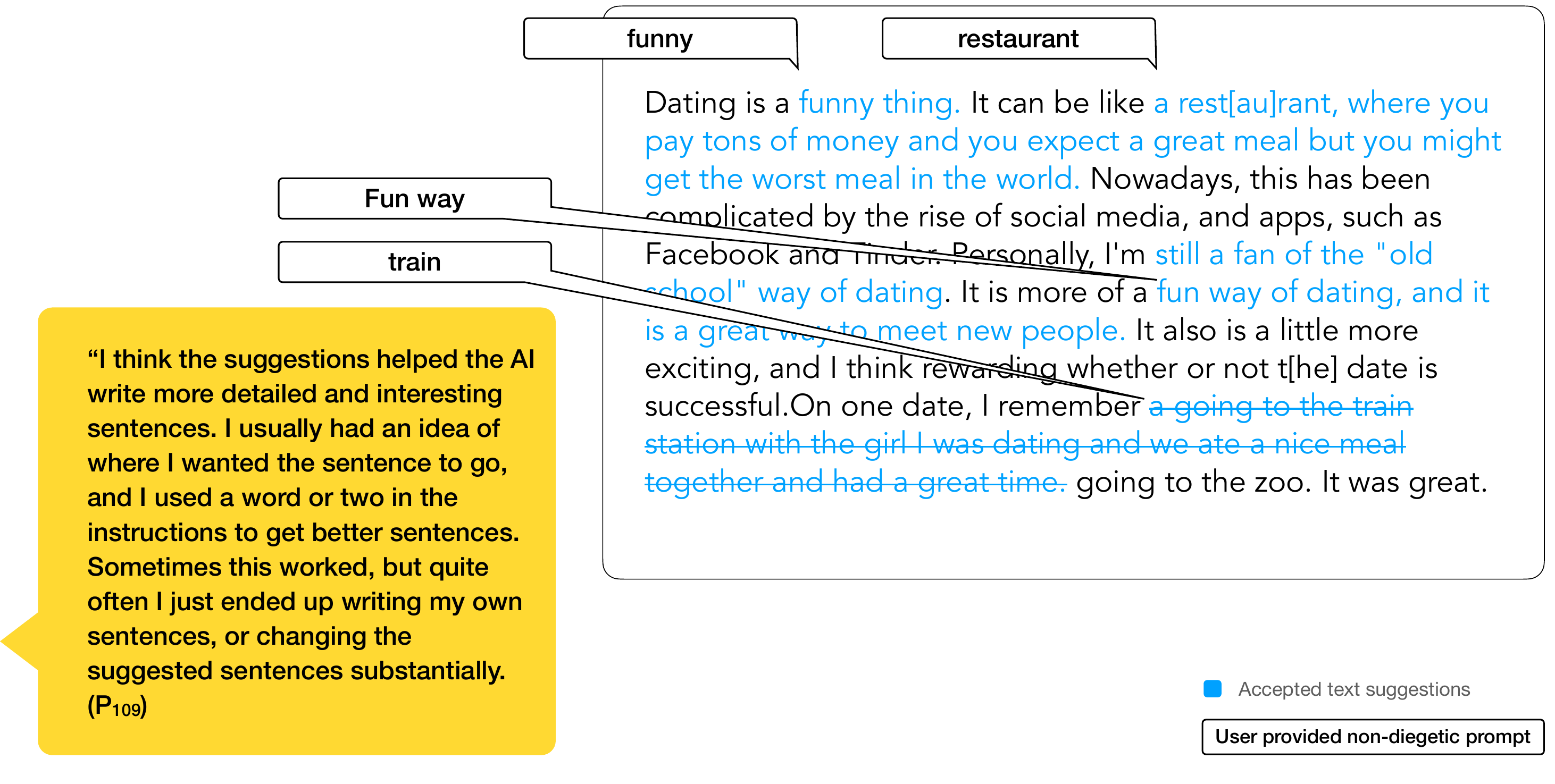}
    \caption{\changenote{Text sample of \participant{109} who provided non-diegetic prompts to guide the LLM. For the last instruction (``train''), \participant{109} decided to then modify the topic to ``zoo''.}}
    \Description{Illustration of one participant who took turns with the AI to write a fictional story}
    \label{fig:example_p109}
\end{figure*}

\changenote{Eighteen participants explicitly commented on suggestion quality: Eight were negative (\participant{91}: \textit{``[...] had to edit most of it.''}).
\changenote{Nine} felt suggestions were hit or miss (\participant{130}: \textit{``Sometimes [the suggestions] helped, sometimes it didn't.''}}). %
\changenote{Two left positive comments} \changenote{(\participant{27}:} \textit{``I was sceptical about whether the AI would align with my ideas or suggest phrasing that I would actually use but most often it did so and I was pleasantly surprised by the results.''}). \changenote{\participant{109} further noted that ``[instructions] helped the AI write more detailed and interesting sentences'' when the direction of the sentence was known beforehand. Using ``one or two words in the instructions to get better sentences'' that participant continued that ``[s]ometimes [it] worked, but quite often I just ended up writing my own sentences, or changing the suggested sentences substantially.'' (see \cref{fig:example_p109}).}

\begin{figure*}
    \centering
    \includegraphics[width=\textwidth]{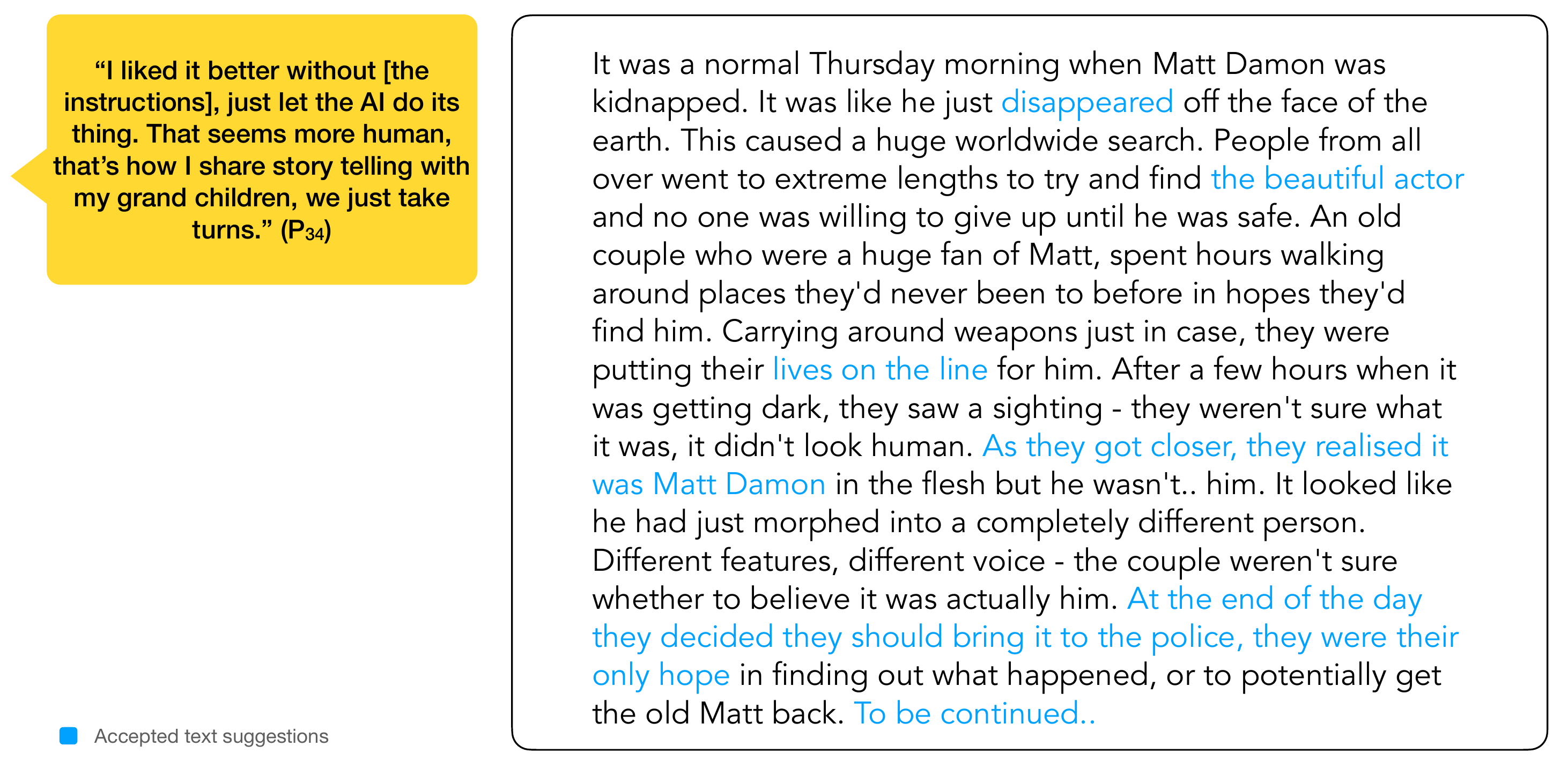}
    \caption{\changenote{Text sample of \participant{34} who took turns with the AI to write about the kidnapping of Matt Damon. The suggestions were taken verbatim and mostly requested at the start or in the middle of a sentence.}}
    \Description{Illustration of one participant who took turns with the AI to write a fictional story}
    \label{fig:example_p34}
\end{figure*}

\subsection{Comments on Instructions}\label{sec:comments_on_instructions}

Opinions diverged on instructions: 21 \changenote{participants} explicitly stated they preferred the UIs allowing for instructions, while 24 preferred those without them. 22 \changenote{participants} reflected on both pros and cons. The main \changenote{reasons for using instructions} were getting more suitable suggestions
\changenote{\coded{12}} %
(e.g. \changenote{\participant{27}:} \textit{``I found the instructions more helpful as I could guide the AI when needed.''}), inspiration for words
\changenote{\coded{29}} %
(e.g. \changenote{\participant{65}:} \textit{``[...] it gave me inspiration when i was stuck for words.''}) and delegating tasks \changenote{like coming up with places, names or synonyms.}
\changenote{\coded{5}}. %
One person \changenote{used the AI} \textit{``[...] to get suggestions for and against the point I was trying to make.'' (\changenote{\participant{85}}}).

In contrast, some found it hard to write instructions (see \cref{sec:open_feedback_learnability} for details). \changenote{Six participants described} a trial-and-error approach to find out how to best write instructions. 

It was also reported that coming up with instructions can disrupt the writing flow \changenote{
\coded{3} %
and thus} reduces efficiency, or is not worth the effort. 
For example: \textit{``It made no difference, as i never felt the need to give it specific instructions. I felt it did a pretty good job of knowing what sort of suggestions I wanted.''} \changenote{(\participant{38})}.

Some said \changenote{writing with} instructions felt less natural \changenote{\coded{3}}: %
\textit{``I mostly enjoyed writing without the instructions. I felt more like I was 'one' with the AI and it felt like it was more of a team member with me than a piece of software. I think because it removed that feeling of using a computer to help me write I felt like the suggested writing was an extension of myself.''} \changenote{(\participant{132})}. And \changenote{\participant{34}} wrote: \textit{``I liked it better without [the instructions], just let the AI do its thing. That seems more human, that's how I share story telling with my grand children, we just take turns.''} \changenote{(see \cref{fig:example_p34}).}

\subsection{Control and Influence}\label{sec:comments_on_control}

\changenote{Eleven participants} commented on control and influencing suggestions. For example: \textit{``I prefer[r]red multiple because - literally - there were multiple to choose from and that gave me a better feeling of control over the story.''} \changenote{(\participant{59})}. Another commented: \textit{``I like the suggestion systems especially when I was able to provide guidance.''} \changenote{(\participant{82})}. Overall, multiple suggestions and instructions were mentioned here as contributing to feeling in control, matching the Likert results on control and influence (\qx{9} and \qx{10} in \cref{fig:likert_agreements}).

Moreover,  \changenote{participants} commented on strategies around what we now call diegetic prompting in this paper. For example, some preferred influencing the suggestion with the diegetic approach: \textit{``I didn't have much success providing instructions, was having trouble thinking of suggestions quickly and instead focused on directing the topic towards a place were viable suggestions would be made without interactive input.''} \changenote{(\participant{99})}. %
Similarly, \changenote{\participant{111}} said: \textit{``Often it was just as difficult to think of the instruction as it would be to actually write something. It seemed just as easy to start writing what I wanted in order to push the AI in the direction I wanted it to go.''}

\begin{figure*}
    \centering
    \includegraphics[width=\textwidth]{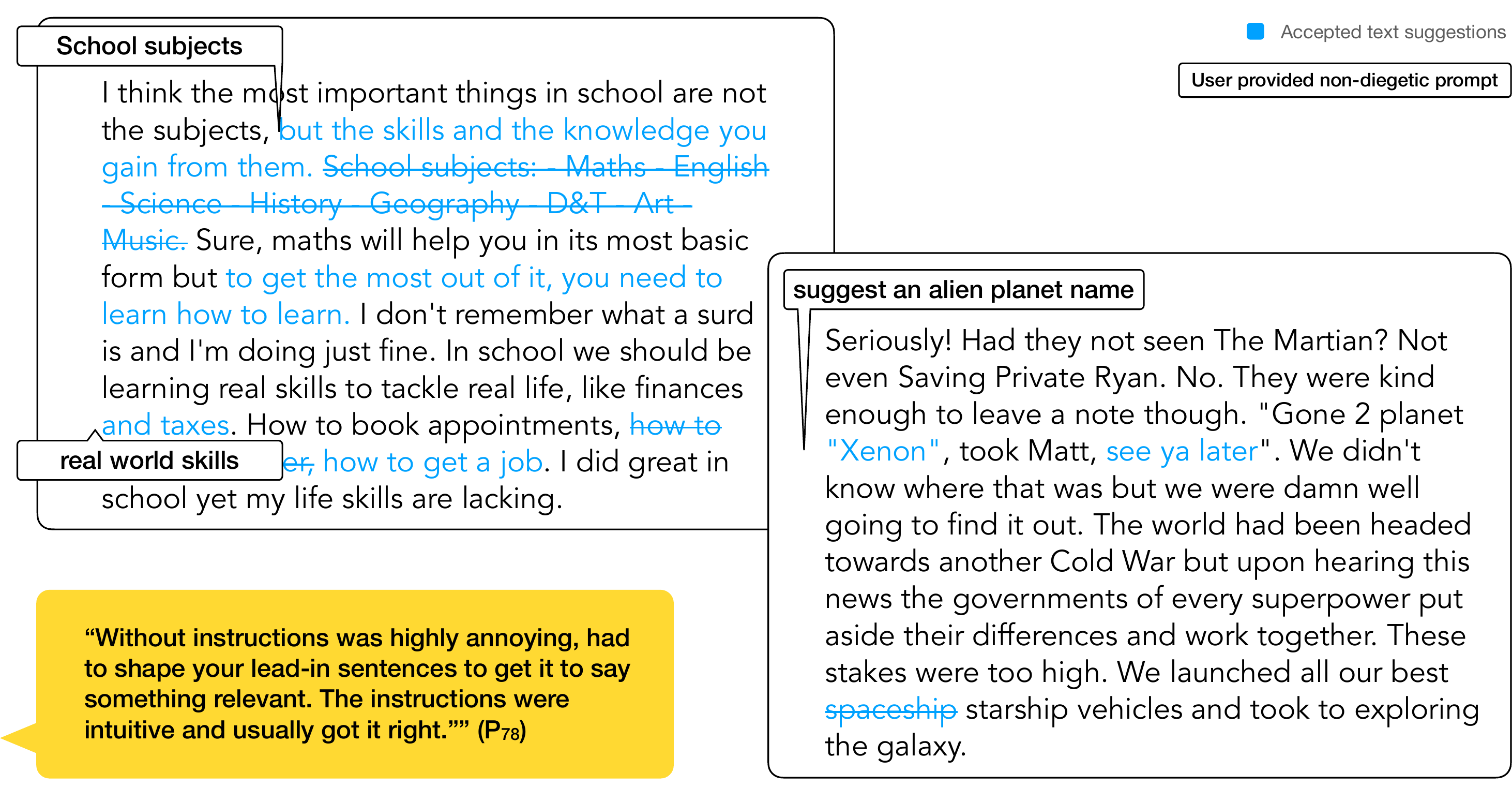}
    \caption{\changenote{Text sample of \participant{78} who used non-diegetic prompting to retrieve a list of ``school subjects''. The accepted suggestion is highlighted in blue. Part of the accepted suggestions was later on deleted.}}
    \Description{Schematic illustration of how one study participant used a non-diegetic prompt to guide the suggested content.}
    \label{fig:example_p78_with_instruction}
\end{figure*}

In contrast, some disliked diegetic prompting: \textit{``Without instructions was highly annoying, had to shape your lead-in sentences to get it to say something relevant. The instructions were intuitive and usually got it right.''} \changenote{(\participant{78}, also see \cref{fig:example_p78_with_instruction})}.

\begin{figure*}
    \centering
    \includegraphics[width=\textwidth]{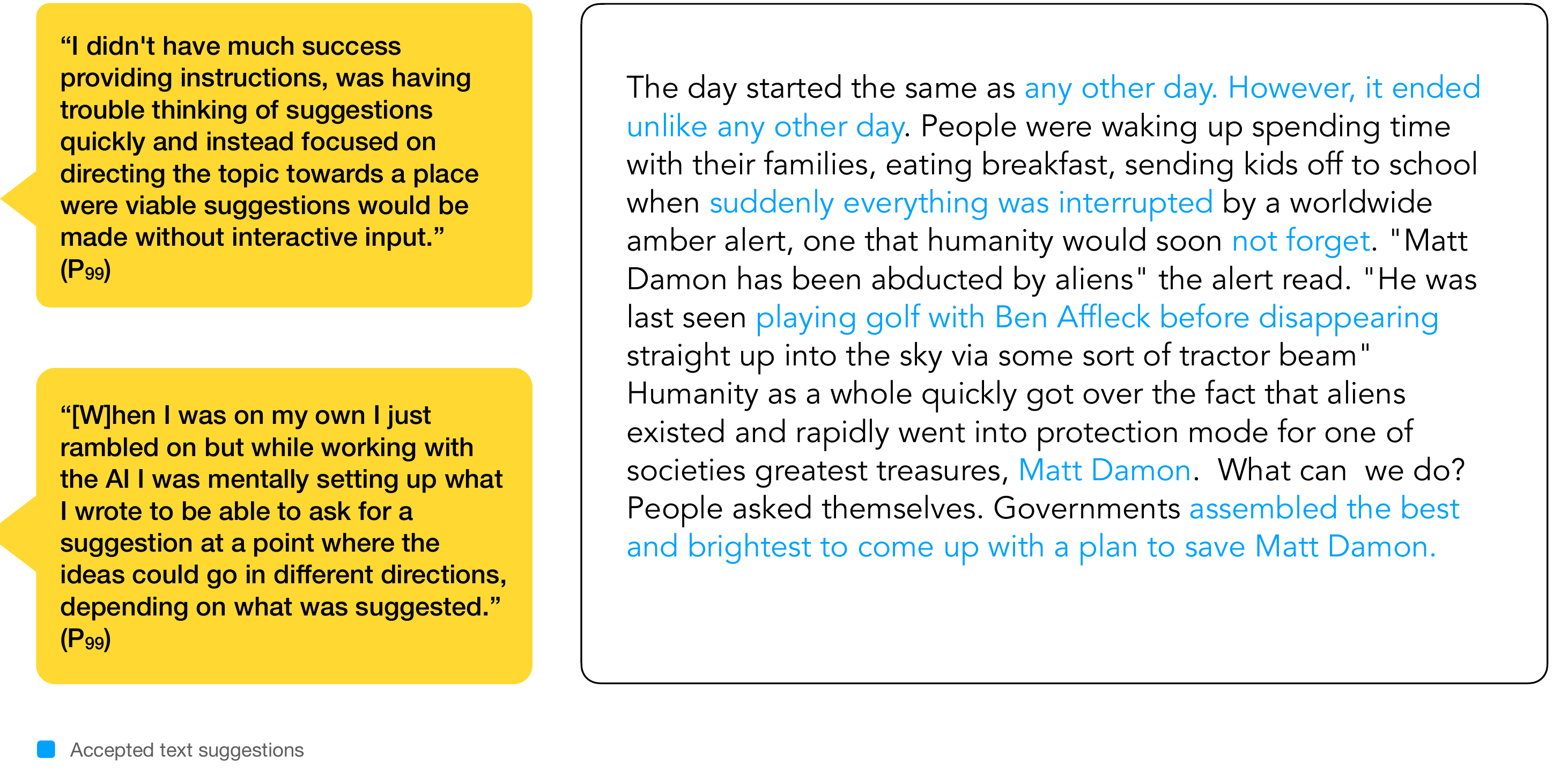}
    \caption{\changenote{Text sample of \participant{99} who found it difficult to provide non-diegetic prompts and instead focused on guiding the suggestion through diegetic content, e.g. requesting suggestions after ``...sending kids off to school when'' (line 3). By setting the sentence up in this way before requesting a suggestion this participant guided the LLM to suggestions that ``go in different directions'' (cf. comment in the second yellow box).}}
    \Description{A schematic illustration of a participant who extended accepted suggestions with their own writing. On the left, there are two quotes describing how that participant struggled to come up with non-diegetic prompts.}
    \label{fig:example_p99_trouble_instructions}
\end{figure*}

Finally, others noticed influences on their own writing processes related to diegetic prompting:
\textit{``[W]hen I was on my own I just rambled on but while working with the AI I was mentally setting up what I wrote to be able to ask for a suggestion at a point where the ideas could go in different directions, depending on what was suggested.''} \changenote{(\participant{99}, also see \cref{fig:example_p99_trouble_instructions})}.
And similarly, \changenote{\participant{9}} wrote: \textit{``[I] noticed that the more time I spent the more my tendency was to find a way to write that would facilitate the suggestion to be meaningful and at the same time interesting to add to give more in-depth to my story.''}

\subsection{Learnability}\label{sec:open_feedback_learnability}

Several \changenote{participants (33)} touched on challenges of learnability and writing instructions: \textit{``I found coming up with suggestions [to the AI] difficult, really. Having to type the start of a sentence and then type what I wanted in a smaller box felt quite clunky and not worth the effort for what was generated. It felt much more fluid when the AI recognized what I wanted and completed the writing without needing suggestions.''} \changenote{(\participant{29})}.
\changenote{\participant{6}} said: \textit{''I almost felt stressed trying to think of some instructions to give to the AI; it felt really hard to me. I'm glad that the option was there, but I guess I wasn't taking full advantage of it.''}
Fittingly, \changenote{25 participants said} they did not use instructions much because, for example, \textit{``[...] I wasn't very good of thinking of them.''} \changenote{(\participant{2})}. 
Some of the previous comments (\cref{sec:comments_on_control}) fit this aspect as well.

\subsection{Distraction}\label{sec:comments_on_distraction}

\changenote{Nine participants} %
explicitly reflected on distraction. For example \changenote{\participant{20} wrote}: \textit{``I actually found the suggestions fairly distracting and not helpful. I tended to already know what I wanted to say so the chances of the suggestions aligning with my thoughts were fairly slim.''} \changenote{\participant{43}} perceived instructions in particular as distracting: \textit{``I feel like writing without instructions help me focus more and [I] am less distracted which allows my sentences to flow and be more natural. Instructions are good if [I] am stuck and need help.''}.

\subsection{Perception of the AI and Expectations}

The comments indicate two fundamental views on the role of the AI: Some expected the system to serve \textit{efficiency}. For example: \textit{``I think there is a lot of value in this system, but inputting instructions make it quite long winded and onerous, negating any benefits there may be. I preferred the multi suggestions without instruction.''} \changenote{(\participant{26})}. 
Also see the first quote on ``alignment'' above (\cref{sec:comments_on_distraction}).
In contrast, others saw the system as serving \textit{inspiration} (also see \cref{sec:comments_on_instructions} and \cref{sec:instruction_usage}). They asked the AI for content suggestions or were curious to see in which direction the AI would take the story.
This included feeling inspired by suggestions even without accepting them: \textit{``I was reading the suggestions either to use them or to just get ideas of what I was writing about''} \changenote{(\participant{67})}.
\section{Discussion}

\subsection{Choice vs. Control}\label{sec:choice_control}

Our findings contribute to the literature on prompt-based interaction with generative systems for writing: %
\changenote{Participants} overall preferred choosing from multiple text suggestions presented to them, over actively writing instructions, \changenote{in short creative and argumentative writing tasks}. This is evident from highest acceptance rates with the multiple suggestions UI (\cref{sec:acceptance_rates}), which were not improved through instructions, and from the qualitative feedback, where a clear majority favored multiple suggestions, while opinions were divided on instructions (\cref{sec:open_comments}). 

However, giving users more control options in the UI by adding non-diegetic prompting partially mitigated the drawback of a lack of suggestion choice: Instructions increased acceptance rates for single suggestions -- although these still did not reach the rate for multiple suggestions (\cref{sec:acceptance_rates}). This indicates that the control offered by instructions was useful to guide single suggestions but not better than having a choice of three suggestions to begin with.

We discuss possible reasons: First, \changenote{participants} might \textit{satisfice}~\cite{Simon1996}, that is, accept a ``good enough'' suggestion rather than trying to ``optimize'' it via instructions. Suggestions might also already be good enough so that there is no need for instructions, as supported by some comments (\cref{sec:comments_on_instructions}).
Second, a known usability principle is \textit{recognition over recall}~\cite{Nielsen1994}: Users might find it easier to recognize a presented suggestion as suitable (or not), compared to coming up with an instruction and typing it in.
Third, \textit{convenience} might lead \changenote{participants in the study} to accept suggestions without instructions to get through the \changenote{tasks} quickly. \changenote{However, participants accepted suggestions at a rate comparable with related work (with multiple suggestions and explicit request via tab key: \pct{74} here vs \pct{72} in \cite{Lee_coauthor_2022}). For suggestions based on user instructions, our rate (\pct{64}) is higher than in a related study design where users could enter requests in a sidebar (\pct{17.6} in \cite{Yuan_wordcraft_2022}): This suggests that potential influences of the study setup do not necessarily work against instructions, or are less dominant than the effects of the UI design (e.g. sidebar vs integration at text cursor). Moreover,} times and texts, in combination with the comments, \changenote{further} support the conclusion that participants took the tasks seriously \changenote{(see \cref{sec:text_evaluation})}. %

At the same time, instructions were indeed (situationally) useful: \changenote{Participants} commented on their benefits (\cref{sec:comments_on_instructions}), used them in every fifth suggestion request, and experimented with different styles (\cref{sec:instruction_usage}). Together, these findings motivate the HCI community to further explore the \textit{integration} of choice and control via prompting. For example, future work could build on our conceptual lens to envision further UI designs that combine diegetic and non-diegetic prompting, and use our data as a benchmark in their evaluation.

\subsection{Guiding Suggestions with Diegetic Prompts}

Our results add to the literature on writing with AI by revealing that people specify more diegetic information to offset the lack of suggestion choice in UIs that display only a single suggestion. 
This is based on the first large-scale analysis of where in the text users request suggestions: 
Users wrote about 1.5 more words in the sentence before requesting single suggestions, compared to multiple ones. 
Moreover, single suggestions were requested less frequently to start a new sentece and to continue after a transition word. Possibly, receiving a single suggestion is less useful here, given that new sentences and transition words signal ``openness'' for potential changes to the direction of the narrative. 

Currently, there is one other (small-scale) analysis of trigger moments (N=4 in \cite{calderwood_how_2020}). Thus, we encourage the community to analyze trigger moments whenever studying UIs with explicit suggestion triggers. %

Fittingly, we indeed recently see high interest in interaction designs where users explicitly request suggestions (e.g. \cite{calderwood_how_2020, Lee_coauthor_2022, Singh_2022}). Our  study explores this design space further by looking at how it interacts with the number of suggestions: Here, we contribute evidence that people consider when to request suggestions, and in particular for single suggestions they request them at points in their text that are expected to give clearer guidance to the text continuation system.
Future work could examine whether this holds in other writing contexts and to what extent users actively think about when to request suggestions while writing. Based on people's comments, at least some strategically thought about what we term diegetic prompting (see \cref{sec:comments_on_control}).

As a related aspect, prior work focused on how people \textit{react} to suggestions (e.g. evaluation fatigue~\cite{Bhat2022}, integrative leaps~\cite{Singh_2022}). Complementary, the above results indicate that there is also a \textit{proactive} direction: Writers think about suggestions \textit{before} seeing them. Future work could investigate this in more detail, in particular for UIs in which users explicitly request suggestions.

\subsection{Challenges of Integrating Non-Diegetic Prompts}\label{sec:learnability_custom_prompts}
We extract two concrete challenges of interacting via non-diegetic prompts to guide future research and design.

\subsubsection{Non-Diegetic Prompts Interrupt the Writing Process}
Writing involves multiple cognitive processes, such as coming up with a thought, turning it into words, and entering it~\cite{hayes_modeling_2012}. Recently, \citet{Bhat2022} studied (without non-diegetic prompts) how this is impacted by text suggestions. For example, writers need to evaluate displayed suggestions. Here, our study adds insights into the relative impact of diegetic vs non-diegetic prompts: Crucially, switching from diegetic writing to non-diegetic instructing forces writers to shift from thinking about their narrative or argument to thinking about instructions to the system. This is reflected in people's comments (\cref{sec:comments_on_instructions}, \ref{sec:comments_on_control}, \ref{sec:open_feedback_learnability}) and the Likert results on distraction and problems with thinking of instructions (\qx{2} and \qx{15} in \cref{fig:likert_agreements}). In contrast, diegetic prompts do not require such shifts, although they still require engagement with displayed suggestions~\cite{Bhat2022, buschek_2021}. 

\subsubsection{Non-Diegetic Prompts can be Hard to Write}
Even after making that shift, then writing effective non-diegetic prompts is difficult, adding to related findings in the literature~\cite{Yuan_wordcraft_2022}: Many participants struggled with this and recognised that they did so in self-reflection (\cref{sec:comments_on_instructions}, \ref{sec:comments_on_control}, \ref{sec:open_feedback_learnability}). More positively, the non-diegetic prompts collected in our study show how users experimented with different styles. These might evolve further with longer use. At the moment, none of these styles go beyond what would also be a meaningful comment to a human co-author.

\subsection{Perceived Role of the AI}\label{sec:perception}

Here we discuss how users perceived the AI and support this discussion by reflecting on three writing processes as in the framework for analyzing writer-suggestion interactions by \citet{Bhat2022}: %
(1) \textit{proposing} new topics or ideas, (2) \textit{translating} abstract thoughts or keywords into sentences, (3) \textit{transcribing} (i.e. entering) words.

\subsubsection{Two Perspectives on the Main Role: Proposer vs Transcriber}
Some people clearly saw the system as something that serves input efficiency (i.e. \textit{transcriber}), whereas others saw it as providing inspiration (i.e. \textit{proposer}). The former are more critical about the system since it would only be good if it is fast and predicts exactly what they want. Based on the qualititative feedback we think that the chosen topic as well as participants' familiarity with the topic might have an influence on their writing mindset. For argumentative writing and, more generally, when people already had an opinion about a topic, they felt that the AI was distracting if it proposed something other than what participants had in mind. Future work may have a closer look at the influence of topic genre and prior knowledge about a topic on the perception of the role of the AI.
Study designs should take this difference into account when choosing writing topics to calibrate metrics for performance or exploration.

\subsubsection{Non-diegetic Prompts Reflect Users' Perception of the AI}

We can further discuss how the content of non-diegetic prompts reflects varying perceptions of the role of the AI:
Considering the writing processes~\cite{Bhat2022}, non-diegetic prompts from our dataset show that users requested the AI to \textit{propose} inspirational ideas. Sometimes users also only provided partial phrases or keywords, or asked for word choices, which puts the AI into the role of \textit{translating} these abstract ideas into full sentences. At other times, they perceived the AI as a \textit{transcriber} for input efficiency (\cref{sec:open_comments}).

Other non-diegetic prompts indicate influences on the perceived role beyond these writing processes: For example, people asked the AI for an opinon or advice, or to lookup information. Thus, non-diegetic prompts may shift perception of the AI's role towards a writing collaborator. %

\subsection{Limitiations and Reflections on Methodology}

People wrote for five minutes with each UI. Hence, they spent ten minutes in total with each individual UI feature across the writing tasks  (single and multiple suggestions, with and without instructions). This is comparable to related work (e.g. 11\,min \cite{Lee_coauthor_2022}, 4\,min \cite{buschek_2021}, 10-12\,min \cite{Yuan_wordcraft_2022}). Future studies should investigate long-term use, in particular to observe how non-diegetic prompts evolve as writers gain experience with a system.

We prototyped our system with GPT-3 via an API. We did not have access to the model directly and we do not claim to have identified the ``best'' settings for our specific usage of the model. We noticed two limitations: Sometimes, suggestions were repetive (e.g. similar ones in one list) or repeated the instruction text (which seems unhelpful). Nevertheless, suggestions were rated highly overall (\cref{sec:results_likert}). %

Potential changes to the model over time are beyond our control. This limits exact replicability for studies like this. We see a trend of limited direct access to state-of-the-art LLMs for parts of the academic community, which is not easy to resolve. On the positive side, our work shows that it is possible to construct and study in detail interactive applications built on existing models.

We chose an online setup in line with recent related work (e.g.~\cite{buschek_2021, Lee_coauthor_2022}) to collect logging data from interactions of many people. However, we could not observe people directly or ask questions at interesting moments in the interaction, except for in our  \changenote{pre-study}, which we used to refine our design. A small-N study with direct observation and think-aloud could complement our work, for example, to understand decision-making around triggering suggestions and writing non-diegetic prompts in more detail. Nevertheless, we received rich qualitative feedback as well (\cref{sec:open_comments}).

It is possible that the instruction styles (\cref{sec:instruction_usage}) are biased by the provided examples (\cref{fig:screenshot_system} in \cref{sec:appendix}). Our pre-study showed that such examples are needed to help people get started with this new feature. Nevertheless, people experimented beyond these examples (e.g. questions, writing help, advice, etc.; see \cref{tab:instruction_examples}). 

\changenote{With the pop-up box, we tested one way of integrating instructions. This UI element is motivated as a simple way of integrating instructions with the established design of a suggestion list (or inline suggestion). A similar pop-up is used in recent related work (not for instructions but for suggestions in the middle of sentences; cf.~\cite{Bhat2022}). Other designs should be explored in the future.}

Finally, we emphasize the importance of open writing tasks in HCI research. Historically, transcription tasks have dominated text entry research (cf.~\cite{Vertanen2014}). With the rising interest in human-AI co-creation, research on writing tools needs new tasks. These might not necessarily focus on measuring input speed but rather cover a range of topics, text types, and other aspects. Pragmatically, writing tasks from writer communities and custom tasks have been used in recent studies (e.g.~\cite{buschek_2021, Lee_coauthor_2022, Singh_2022, Yuan_wordcraft_2022}), including ours. As a community, we should systematically evaluate and curate such writing tasks if they are to become a lasting key methodological component.

\subsection{Beyond Writing: Diegetic and Non-diegetic Interaction in Generative Systems}

We have studied diegetic and non-diegetic prompts to draft text (i.e. \textit{text to text}). Here we reflect on this new perspective by discussing concrete examples of how other interactive generative systems use diegetic and non-diegetic prompting.

\begin{itemize}

\item{\textit{Visual to Text}}
\citet{chung_talebrush_2022} proposed a new story ideation tool that uses visual sketching to guide a LLM. Here the sketch is translated to a text prompt. This interaction is non-diegetic.

\item{\textit{Text to Visual}}
Recent text to image models allow users to generate images from text descriptions \cite{patashnik_styleclip_2021, nichol_glide_2021, ramesh_hierarchical_2022}. These are non-diegetic prompts, because they are not part of the visuals.

\item{\textit{Visual to Visual}}
\citet{bau_gan_2018} show an example of ``painting shapes'' to guide image models: Users draw simple shapes such as a triangle to symbolize a mountain. The image model then translates these shapes into a high-fidelity rendering. Since the abstract shape is usually not part of the outcome we consider this interaction non-diegetic. On the other hand, \citet{ha_neural_2017} enable users to start painting a part of an image (i.e. providing diegetic information) and let the system continue or finish the painting.
\end{itemize}

Differentiating these two perspectives therefore allows researchers to analyse users' intention and behavior when interacting or designing systems with generative AI. As shown in the following discussion we can use this understanding to derive implications on the design of interactions for generative models.

\subsection{Implications for LLMs and User Interfaces}\label{sec:implications}
In recent work by \citet{schick_peer_2022}, their LLM ``PEER'' is explicitly trained to follow non-diegetic prompts related to text revision. Effectively, our study contributes the HCI counterpart -- an investigation of a UI and interaction design to integrate an LLM in such a role into the writing process. Our results guide future work at this intersection of HCI and NLP in two concrete ways: 

First, based on our collected non-diegetic prompts these LLMs should be trained to understand a broader range of inputs. For instance, PEER is trained on the \textit{imperative}-style but we found the \textit{keyword}-style to be more common. Alternatively, users need to be guided towards the supported style via the UI. 

Second, while LLMs are rapidly improving, even the best model cannot eliminate cognitive costs and interaction costs of switching between diegetic and non-diegetic writing. This motivates further studies on interaction designs that require such switches and potential pathways to making them easier and more efficient.

\section{Conclusion}

Our new understanding highlights that people use two types of prompting to guide LLMs for text generation. While related work has presented systems that focused on non-diegetic prompts, our findings reveal that users additionally think about and shape their text to guide LLMs through diegetic information.
\changenote{With our UI design that allows for both types, using GPT-3, participants} preferred choosing from multiple suggestions over writing instructions.
We conclude by highlighting three key takeaways based on our results:

First, writing instructions to the AI requires effort, including switching between diegetic and non-diegetic writing. %
Second, people combine diegetic and non-diegetic prompting, as single suggestions benefitted from both. %
Third, writers use their draft (i.e. diegetic information) and suggestion timing to strategically guide LLMs, based on our analysis of when people request suggestions, as well as their self-reflection in comments. %

We encourage future work to further analyze these prompt types to develop better writing tools and generalize to other domains (e.g. interaction with generative models for images). To facilitate this, we release our prototype and material on the study and analysis here:

\url{https://osf.io/qwakj}

\begin{acks}
We thank Lukas Mecke for feedback on the manuscript. This project is funded by the Bavarian State Ministry of Science and the Arts and coordinated by the Bavarian Research Institute for Digital Transformation (bidt).
\end{acks}

\bibliographystyle{ACM-Reference-Format}
\bibliography{references}

\appendix
\section{Appendix}\label{sec:appendix}
In this appendix, we provide a table of logged events and additional screenshots.

\begin{table*}[h]
\centering
\footnotesize
\newcolumntype{L}{>{\raggedright\arraybackslash}X}
\newcolumntype{P}[1]{>{\raggedright\arraybackslash}p{#1}}
\renewcommand{\arraystretch}{1.4}
\setlength{\tabcolsep}{4pt}
\begin{tabularx}{\linewidth}{lP{20em}L}

\toprule
No.            
    & 
    Interaction Event                                        
    & 
    Description \\ \midrule
1  & 
    EVENT\_CONFIRM\_INSTRUCTION       
    & 
    User has confirmed instruction (Enter Key) \\
2  & 
    EVENT\_CANCEL\_INSTRUCTION        
    & 
    User has cancelled the instruction (ESC key or clicking outside the instruction box) \\
3  & 
    EVENT\_OPEN\_INSTRUCTION\_BOX     
    & 
    User has triggered new suggestions in the ``with instructions'' writing setting (Tab Key) \\
4  & 
    EVENT\_SELECT\_NEXT\_SUGGESTION   
    & 
    User has selected next suggestion (Down Arrow Key) \\
5 &
    EVENT\_SELECT\_PREV\_SUGGESTION   
    & 
    User has selected previous suggestions (Up Arrow Key) \\ 
6  & 
    EVENT\_REQUEST\_SUGGESTIONS       
    & 
    User has requested new suggestions (Tab Key) \\
7  & 
    EVENT\_SUGGESTIONS\_RESPONSE      
    &
    System returned suggestions \\
8  & 
    EVENT\_CONFIRM\_SUGGESTION        
    & 
    User has selected and confirmed one suggestion (Enter Key or Mouse Selection) \\
9 & 
    EVENT\_CANCEL\_SUGGESTION         
    & 
    User has cancelled the suggestions (ESC key or clicking outside the suggestion box) \\
10  & 
    EVENT\_TASK\_STATUS               
    & 
    Can be either ``task started'' or ``task finished'' \\
11 & 
    EVENT\_KEYDOWN   
    & 
    User has pressed a key, e.g ``A'' or ``TAB'' \\ 
    \bottomrule
\end{tabularx}
\label{tab:event_log}
\caption{An overview of the interaction events logged in the user study.}
\end{table*}

\begin{figure*}
    \centering
    \includegraphics[width=\textwidth]{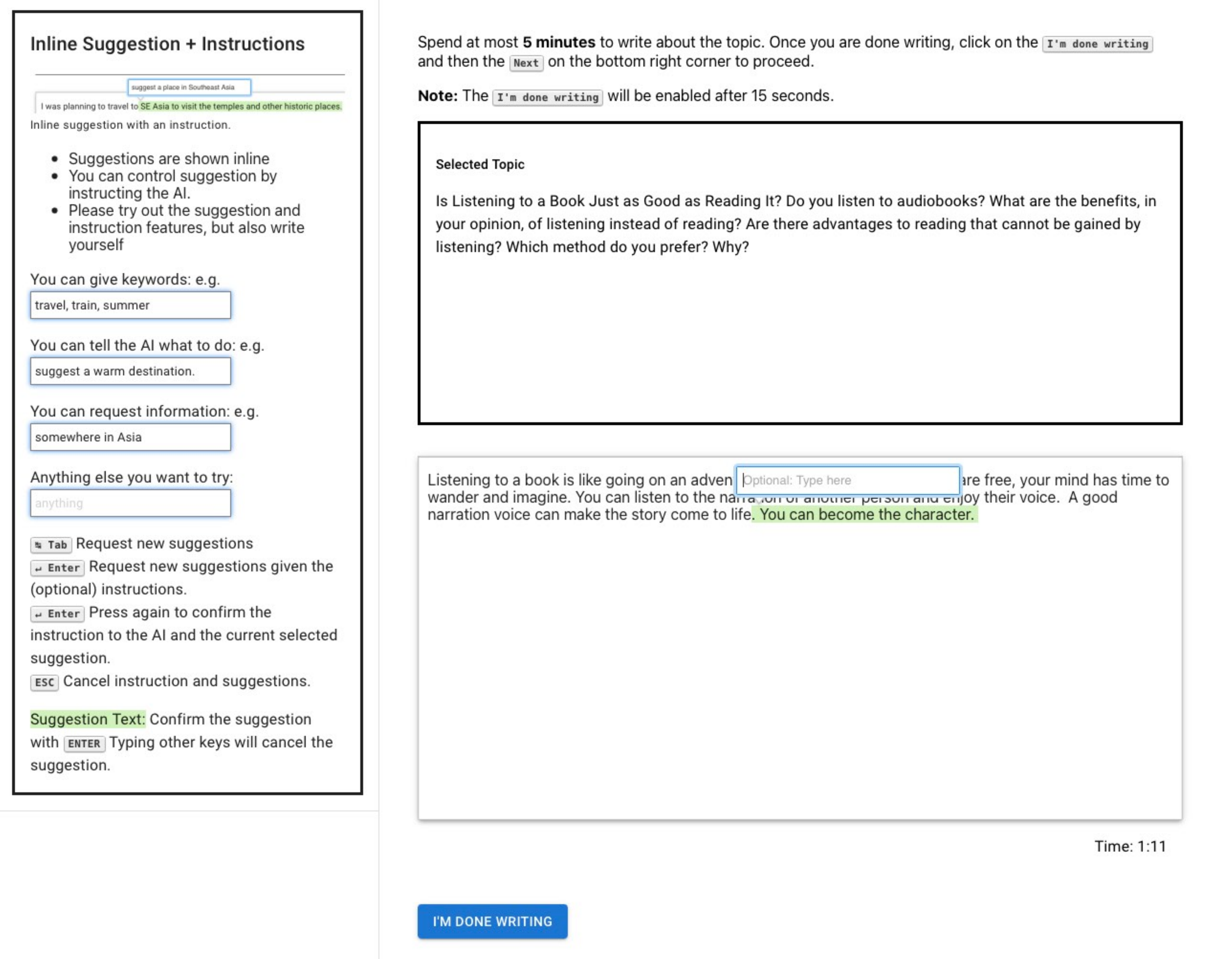}
        \caption{Screenshot of the writing interface. (Left Side) The info box describes the available functionalities in the current setting, (Top Middle) the selected topic, (Bottom Middle) the text editor with the current written text and an inline suggestion.}
        \Description{Screenshot of the writing interface after participants have chosen a topic to write about.}
    \label{fig:screenshot_system}
\end{figure*}

\begin{figure*}
    \centering
    \includegraphics[height=8cm]{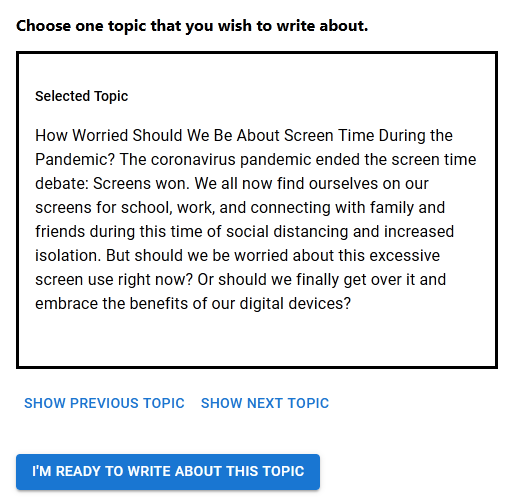}
    \caption{\changenote{The topic selection panel. Users can browse through the topics and indicate that they are ready to write about the depicted topic.}}
    \Description{Screenshot of the topic selection panel.}
    \label{fig:topic_selection}
\end{figure*}

\end{document}